\documentclass[aps,prl,reprint,showpacs,amsmath,amssymb,superscriptaddress,floatfix]{revtex4-1}
\usepackage{amsmath}
\usepackage{amssymb}
\usepackage{bm}
\usepackage{graphics}
\usepackage{float}
\usepackage{color}
\usepackage{bbold}
\usepackage{graphicx}
\usepackage{ifpdf}
\usepackage{epstopdf}
\usepackage{bbold}

\begin{document}
\title{Stability of a topological insulator: interactions, disorder and parity of Kramers doublets}

\affiliation{Institute of Fundamental and Frontier Sciences, University of Electronic Science and Technology of China, Chengdu 610054, People's Republic of China}
\affiliation{Shamoon College of Engineering, Bialik/Basel St., Beer-Sheva 84100, Israel}
\affiliation{Institut f\"ur Theoretische Physik, Universit\"at Hamburg,  
	Jungiuustr.9, D-20355 Hamburg, Germany}
\affiliation{Aston University, School of Engineering \& Applied Science, Birmingham B4 7ET, UK}

\author{V. Kagalovsky}
\affiliation{Institute of Fundamental and Frontier Sciences, University of Electronic Science and Technology of China, Chengdu 610054, People's Republic of China}
\affiliation{Shamoon College of Engineering, Bialik/Basel St., Beer-Sheva 84100, Israel}
\author{A. L. Chudnovskiy}
\affiliation{Institut f\"ur Theoretische Physik, Universit\"at Hamburg,  
	Jungiuustr.9, D-20355 Hamburg, Germany}
\author{I. V. Yurkevich}
\affiliation{Institute of Fundamental and Frontier Sciences, University of Electronic Science and Technology of China, Chengdu 610054, People's Republic of China}
\affiliation{Aston University, School of Engineering \& Applied Science, Birmingham B4 7ET, UK}

\begin{abstract}
We study stability of multiple conducting edge states in a topological insulator against all multi-particle perturbations allowed by the time-reversal symmetry. We model a system as a multi-channel Luttinger liquid, where the number of channels equals the number of Kramers doublets at the edge. We show that in the clean system with $N$ Kramers doublets there {\it always} exist relevant perturbations (either of superconducting or charge density wave character) which {\it always} open $N-1$ gaps. In the charge density wave regime, $N-1$ edge states get localised. The single remaining gapless mode describes sliding of 'Wigner crystal' like structure. Disorder introduces multi-particle backscattering processes. While the single-particle backscattering turns out to be irrelevant, the two-particle process may localise this gapless, in translation invariant system, mode. Our main result is that an interacting system with $N$ Kramers doublets at the edge may be either a trivial insulator or a topological insulator for $N=1\, {\rm or}\, 2$ depending on density-density repulsion parameters whereas any higher number $N>2$ of doublets gets fully localised by the disorder pinning irrespective of the parity issue.
\end{abstract}


\maketitle
%\section{Introduction}

{\it Introduction} -- Topological insulators (TI) have attracted great attention in condensed matter physics \cite{KM, BAB}.
The main feature of 2D topological insulators is the existence of conducting edge states protected by the time-reversal symmetry (TRS). Each edge state is a helical Kramers doublet (KD) with opposite spins propagating in opposite directions. TRS forbids a spin-flip backscattering within the same KD, but allows it between two different KDs. In a non-interacting system, a backscattering between different doublets generated by a disorder localises all edge states for even number of KDs and allows odd number (at least one) of delocalised edge modes if the number of KDs is odd \cite{Bardarson}. The former case then corresponds to a trivial insulator whereas the latter must be referred to as topological. The TRS argument \cite{Bardarson} then states that the main distinction between topological and trivial insulators is the parity of the number of Kramers doublets. This is the conclusion reached on the basis of symmetries of the scattering matrix which is valid for non-interacting systems only. The effect of interactions on the edge states behavior under perturbations  is of a great importance. It was studied intensively for a system with a single KD \cite{chu1,chu2,chu3,chu4}, and for systems with one or more KDs  \cite{Moore,San_G,San_G2,neu,ster}.  One of the main conclusions of these studies was that an even number of KDs can be stabilised by interactions and remain conducting. On the other hand, to the best of our knowledge the existing experiments provide so far only evidence of the existence of 2D topological insulators with a single KD \cite{Wurz}. 

In this Letter, we consider an arbitrary number $N$ of KDs existing at the edge of a 2D material. Assuming realistic situation that all KDs exist within a layer which is narrower than a screening radius, we apply model of featureless (Coulomb-blockade or 'orthodox' model) interaction between them. We will show that for generic interaction parameters CDW instability of repulsive fermions in a clean (translation invariant) system leads to the formation of a rigid structure (similar to the Wigner crystal in higher dimensions) stemming from the freezing of $(N-1)$ gapped modes. The remaining single gapless mode describes sliding of the total charge and it gets pinned by a backscattering term generated by a random inhomogeneity leading to a full localisation of the edge modes when number of Kramers doublets exceeds two, $N>2$. The conductance is not fully suppressed by disorder in two situations only. In the case of a single Kramers doublet, $N=1$, no gaps due to interaction could be generated and the dimensionless edge conductance may be equal to one for a wide range of parameters. A pair of doublets, $N=2$, also may survive pinning by disorder (maintaining dimensionless edge conductance equal to two) but the stability region is small and, therefore, difficult to reach and observe experimentally. 

Note that we are interested in the weak interaction problem, hence not all symmetry allowed scattering process create spectral gaps.  Rather, only processes that are relevant in the renormalisation group (RG) sense become potential candidates for opening  gaps in the excitation spectrum. This is in contrast to a strong interaction problem (see, for example \cite{neu}, and references therein), in which case  all symmetry allowed interactions have to be taken into account on equal footing, and the Haldane criterion \cite{hal} must be applied to singling out the maximal number of consistent conditions for spectral gaps.  

We will show below that for $N$ repulsive KDs one can always find $(N-1)$ interaction processes that glue together density profiles of different KDs creating single conducting mode (CDW regime) that may slide in TRS system. There is another region of parameters where CDW gets pinned and TRS is spontaneously broken. For this set of RG-relevant interactions, the Haldane criterion \cite{hal} is automatically satisfied and, therefore, our analysis is insensitive to the parity of KDs number.

The proper way to describe a one-dimensional physics with interactions is the Luttinger liquid (LL) theory \cite{Gim} description. To consider multiple edge states, one has to study a multi-channel system in the framework of sliding Luttinger liquid (sLL) \cite{Sondhi,sLL,Kane2002,smectic,XY}. It is convenient to define a Luttinger matrix $\hat K$ \cite{Yur1, Yur2, KLY, Yur, ACh, JLY}  which is a generalisation of a Luttinger parameter $K$ for a single channel. All scaling dimensions of all symmetry allowed perturbations can be expressed using this single matrix ${\hat K}$. This matrix provides information on the relevance of perturbations and, therefore, a stability region for a topological insulator. 

This manuscript is organised as follows: we start with the formulation of the model and introduction of the perturbations present in a clean (translation invariant) system. The renormalisation group (RG) analysis of this model will allow us to single out gapless modes and formulate low-energy effective model. We will then treat the effect of a disorder on the survived low-energy mode and build the phase diagrams with the focus on robustness of topological insulators against random disorder.

{\it The model} --
The Lagrangian describing a multichannel Luttinger liquid is built on two vector fields,  ${\bm\phi}=(\phi_1\,,...\,,\phi_N)$ and ${\bm\theta}=(\theta_1\,,...\,,\theta_N)$, parametrising excitation densities, $\rho_i=\partial_x\phi_i/2\pi$, and currents, $j_i=\partial_x\theta_i/2\pi$, in each channel $i$ ($1\leq i \leq N$)\cite{Sondhi,sLL,Kane2002,smectic,XY}. The Lagrangian, ${\cal L}_0$, written in terms of the composite field
${\bm\Psi}^{\rm T}=({\bm\phi}^{\rm T}\,,{\bm\theta}^{\rm T})$,
\begin{equation}\label{L0}
{\cal L}_0=\frac{1}{8\pi}{\bm \Psi}^{\rm T}\,\left[{\hat\tau}_1\,\partial_t+{\hat V}\,\partial_x\right]\,\partial_x\,{\bm \Psi}\,,
\end{equation}
includes block-diagonal matrix ${\hat V}={\rm diag}[{\hat V}_+\,,{\hat V}_-]$ with each block describing density-density, ${\hat V}_+$, and current-current, ${\hat V}_-$, interactions; ${\hat\tau}_1$ is the Pauli matrix.
Interaction matrices for KDs in topological insulators should be distinguished from a standard multi-channel (array of wires) model, where inter-wire interactions decay with the distance between wires, or even only nearest-neighbour (nearest, adjacent wires) interaction is assumed. Since all KDs are localised near an edge there spatial separation can be much shorter than the screening length of the interaction. This is the model we analyse below. Taking all velocities equal each other and we can put them equal unity. Inter-KDs interactions are assumed to be equivalent for all KDs,:
\begin{equation}
V^{ij}_{\pm}=\left(1+g_{\pm}\right)\,\delta_{ij}+g'_{\pm}\,\left(1-\delta_{ij}\right)\,,
\end{equation}
All parameters are defined following standard nomenclature: $g_{\pm}=g_4\pm g_2$ with coupling $g_4$ being an interaction strength between  electrons moving in the same direction (right- with right-movers, and left- with left-movers), and $g_2$ is the interaction strength between electrons moving in the opposite directions within the same KD. The couplings with prime have similar meaning for inter-channel interactions.

It is convenient to represent the matrices as sums of two terms acting in orthogonal subspaces,
\begin{equation}\label{V}
\hat{V}_{\pm}=v_{\parallel}\,K_{\parallel}^{\pm 1}\,{\hat\Pi}+v_{\perp}\,K_{\perp}^{\pm 1}\left(\hat{\mathbb{1}}-{\hat\Pi}\right)\,,
\end{equation}
with two projectors in channel space defined by
\begin{equation}
\hat{\Pi}=N^{-1}\,{\bf e}\otimes{\bf e}\,,\quad {\bf e}=\left(1,1, ..., 1\right)\,,\quad
{\hat \Pi}_{\perp}={\hat{\mathbb{1}}}-{\hat\Pi}.
\end{equation}
The 'effective' Luttinger parameters
\begin{equation}\label{K}
K_{\perp}=K\,\sqrt{\frac{1-\alpha_-}{1-\alpha_+}}\,,\quad K_{\parallel}=K\,\sqrt{\frac{1+(N-1)\alpha_-}{1+(N-1)\alpha_+}}\,,
\end{equation}
are related to to the standard Luttinger parameter $K$ defined in the absence of inter-channel interactions and inter-channel couplings $\alpha_{\pm}=g'_{\pm}/(1+g_{\pm})$ (we omit definitions of the velocities because their values are irrelevant for the analysis below).

{\it Interactions} - The model of interacting KDs contains terms describing multi-particle interactions beyond (forward scattering) quadratic Lagrangian. The most general interaction is written as
\begin{equation}\label{int}
{\cal L}_{\rm int}=\sum\limits_{Q=0}\,h({\bf j},{\bf q})\,e^{i({\bf j}{\bm\phi}+{\bf q}{\bm\theta})}\,,
\end{equation}
where the summation is restricted by the neutrality requirement $Q=0$ since the charge of the vertex (number of created minus number of annihilated particles) labelled by pair $({\bf j},{\bf q})$  is equal to $Q=2{\bf q}{\bf e}$. The vectors ${\bf j}$ and ${\bf q}$ have components that take integer and half-integer values and corresponding components, $j_i$ and $q_i$, must be both either integer or half-integer. 

The possible amplitudes of the couplings are related to each other by hermiticity ${\bar h}({\bf j},{\bf q})=h(-{\bf j},-{\bf q})$ and time-reversal symmetry (TRS):
\begin{equation}\label{TRS}
h({\bf j},{\bf q})=h({\bf j},-{\bf q})\,(-1)^J\,,\quad J={\bf j}{\bf e}\,.
\end{equation}
Note that the neutrality requirement $Q=0$ implies that $J$ is an integer.

{\it Relevance of perturbations} -- 
Since we are dealing with a weak interaction case, not all perturbations present in Eq.~(\ref{int}) dictate the system state. Only those terms that are relevant in the renormalisation group sense should be taken into account. Discarding irrelevant terms we will be left with the effective low-energy action. The scaling dimension of an arbitrary vertex from interactions Eq.~(\ref{int}) can be written as
\begin{equation}\label{d}
\Delta({\bf j},{\bf q})={\bf j}\,{\hat K}\,{\bf j}+{\bf q}\,{\hat K}^{-1}\,{\bf q}\,,
\end{equation}
where matrix ${\hat K}$ is the generalisation of the single-channel Luttinger parameter to multi-channel case (please note that it is not a statistics matrix, sometimes called ${\cal K}-matrix$, used in description of fractional liquids). The matrix ${\hat K}$ employed in Eq.~(\ref{d}) is the solution of the algebraic matrix equation \cite{Yur,KLY}:
\begin{equation}\label{K}
{\hat K}\,{\hat V}_+\,{\hat K}={\hat V}_-\,.
\end{equation}
Solving this equation for the interaction matrices ${\hat V}_{\pm}$ defined in the Eq.~(\ref{V}),
\begin{equation}
{\hat K}=K_{\parallel}\,{\hat\Pi}+K_{\perp}\,{\hat\Pi}_{\perp}\,,
\end{equation}
one easily finds the scaling dimensions of the vertices in the interaction term:
\begin{equation}\label{delta}
\Delta({\bf j},{\bf q})=K_{\perp}\,{\bf j}^2+K^{-1}_{\perp}\,{\bf q}^2+(K_{\parallel}-K_{\perp})\,\frac{J^2}{N}\,.
\end{equation}
The perturbations in Eq.~(\ref{int}) may have random amplitudes $h$ with mean zero value (stemming from disorder) and non-random amplitudes allowed in a translation  invariant system. They should be treated differently. Let us first analyse the latter.

{\it Clean system} --
Perturbations allowed in a translation invariant system are further restricted by the momentum conservation $J=0$. The scaling dimensions of zero-current, $J=0$, interactions 
\begin{equation}\label{delta-inv}
\Delta({\bf j},{\bf q})=K_{\perp}\,{\bf j}^2+K^{-1}_{\perp}\,{\bf q}^2\,,
\end{equation}
The most RG dangerous terms are known (see e.g. \cite{stab1,stab2, stab3, stab4, stab5}). They correspond to the minimal values of the scaling dimensions. There are three different terms but one of them that corresponds to the choice ${\bf j}=\pm{\bf q}={\bf t}_{ij}/2$ (where vector ${\bf t}_{ij}=(0,...,1_i, ..., -1_j, ..., 0)$ with arbitrary $i\neq j$) is a single-particle scattering and will be ignored in our analyses because these processes have been accounted for in constructing a non-interacting KDs model. The other two terms are known to be responsible for charge density wave, 
\begin{equation}
{\cal L}^{\rm cdw}\sim\sum\,h^{\rm cdw}_{ij}\,e^{i(\phi_i-\phi_j)}\,,
\end{equation}
with scaling dimension $\Delta^{\rm cdw}=\Delta({\bf t}_{ij}\,,0)$, and superconductivity,
\begin{equation}
{\cal L}^{\rm sc}\sim\sum\,h^{\rm sc}_{ij}\,e^{i(\theta_i-\theta_j)}\,,
\end{equation}
 with scaling dimension $\Delta^{\rm sc}=\Delta(0\,,{\bf t}_{ij}$). The explicit expressions for their scaling dimensions can be found from Eq. \eqref{delta-inv}:
\begin{eqnarray}
\Delta^{\rm cdw}=2\,K_{\perp}\,,\quad \Delta^{\rm sc}=2\,K^{-1}_{\perp}\,.
\end{eqnarray}
Translation invariant perturbations are RG-relevant when their scaling dimensions are below the physical dimension $d$ and $d=2$ in this case. Note that one of two two-particle perturbations is {\it always} relevant (smaller than $2$), and therefore freezes $N-1$ differences between corresponding bosonic fields and opens $N-1$ gaps. Before we turn to the effect a disorder on the remaining single gapless mode, we have to separate gapped and gapless modes to write the effective low-energy field theory of the translation invariant system. This task can be achieved by an orthogonal transformation on both ${\bm\phi}$- and ${\bm\theta}$-vector fields to diagonalise Hamiltonian and preserve commutations. The orthogonal matrix of the form ${\hat O}=({\bf e}_1\,, ..., {\bf e}_{N-1}\,,{\bf e}/\sqrt{N})$ (with all mutually orthogonal vectors) will achieve the goal. Same procedure can be described by the following separation of the vector fields into orthogonal subspaces using projector introduced above,
\begin{equation}
{\bm\phi}=\frac{\Phi\,{\bf e}}{\sqrt{N}}+{\bm\phi}_{\perp}\,,\quad
{\bm\phi}_{\perp}={\hat\Pi}_{\perp}{\bm\phi}\,,
\end{equation}
and similar expression for the conjugate ${\bm\theta}$-fields. This transformation may be thought of as an introduction of the 'centre-of-mass' coordinates ($\Phi$ and $\Theta$) and the relative to it $(N-1)$ 'positions' ${\bm\phi}_{\perp}$ and ${\bm\theta}_{\perp}$. The Lagrangian in the new fields decomposes into two terms ${\cal L}_0={\cal L}_{\perp}+{\cal L}_{\parallel}$. The fields ${\bm\Psi}_{\perp}=({\bm\phi}_{\perp},{\bm\theta}_{\perp})$ are gapped by $(N-1)$ RG-relevant terms
\begin{eqnarray}\nonumber
{\cal L}_{\perp}&=&\frac{1}{8\pi}\,{{\bm\Psi}}_{\perp}\left[\tau_1\partial_t+v_{\perp}
\,{\hat \kappa}_{\perp}\,\partial_x\right]\partial_x{{\bm\Psi}}_{\perp}\\
&+&\sum_{Q=J=0}\,h({\bf j}, {\bf q})\,e^{i({\bf j}{\bm\phi}_{\perp}+{\bf q}{\bm\theta}_{\perp})}
\end{eqnarray}
where ${\hat \kappa}_{\perp}={\rm diag}\left[K^{-1}_{\perp}\,{\hat{\mathbb{1}}}\,,K_{\perp}\,{\hat{\mathbb{1}}}\right]$. 
The 'internal' degrees of freedom are necessarily gapped by either CDW or SC coupling.
For the repulsive interaction, the case we are analysing in this paper, $K_{\perp}<1$ and the most dangerous terms with scaling dimension $\Delta<2$ are $(N-1)$ terms with ${\bf q}=0$ and $J=0$.

The 'centre-of-mass' coordinates drop out of all terms describing inter-channel coupling in a translation invariant system due to $J=0$ restriction. The corresponding mode cannot be gapped and the Lagrangian of gapless $\Phi$ and $\Theta$,
\begin{equation}
{\cal L}_{\parallel}=\frac{1}{4\pi}\,\partial_t\Theta\,\partial_x\Phi-
\frac{v_{\parallel}}{8\pi}\left[\frac{1}{K_{\parallel}}\left(\partial_x\Phi\right)^2
+K_{\parallel}\left(\partial_x\Theta\right)^2\right]\,,
\end{equation}
describes low-energy behaviour.

{\it Disorder} --
Inhomogeneity breaks translation invariance allowing $J\neq 0$ terms to appear in the Hamiltonian. Allowed terms should not contain gapped modes: ${\bm\phi}_{\perp}$-field is frozen and the conjugate to it  ${\bm\theta}_{\perp}$-field would make corresponding terms irrelevant (in particular, single-particle inter-channel backscattering). The field $\Theta$ cannot appear in the interactions due to $Q=0$ neutrality restriction. This consideration leads to the following Lagrangian describing low-energy disordered system of $N$ KDs:
\begin{equation}\label{dis}
{\cal L}_{\rm dis}={\cal L}_{\parallel}+\sum_{n=1}^{\infty}\,h_{2n}\,e^{i\frac{2n}{\sqrt{N}}{\Phi}}\,,\quad
h_J=\sum_{\left\{{\bf j}:J\neq 0\right\}}\,h({\bf j})\,.
\end{equation}
The restriction $J=2n$ in this summation is the result of TRS requirement $(-1)^J=1$ (see Eq.~(\ref{TRS}) with ${\bf q}=0$). The scaling dimension of each interaction term in Eq.~(\ref{dis}) follows from Eq.~(\ref{delta}):
\begin{equation}
\Delta_J=J^2K_{\parallel}/N\,.
\end{equation}
Most dangerous term satisfying TRS corresponds to $J=2$. One of the examples would be 
a simultaneous backscattering of two particles in two different channels,
\begin{equation}
L^{J=2}_{\rm dis}\sim \int\,dx\,\xi_{ij}(x)\,{\bar R}_i\,{\bar R}_j\,L_i\,L_j+\mathrm{c. c.}\,.
\end{equation}
with random anti-symmetric matrix $\xi_{ij}$. Since disorder assumes zero mean value of $\xi_{ij}$, the scaling dimension of this term should be compared with $3/2$.

If $J=2$ backscattering terms (pinning potential for a structure similar to the Wigner crystal) were irrelevant, this single mode would be conducting with dimensionless conductance equal to the total number of the Kramers doublets. The conductance cannot be changed by irrelevant perturbations acting on the collective 'centre-of-mass' coordinate and this fact is reflected in the relationship between total density and current and the centre-of-mass variables:
\begin{equation}
\rho=\frac{\sqrt{N}}{2\pi}\,\partial_x\,\Phi\,,\quad j=\frac{\sqrt{N}}{2\pi}\,\partial_x\,\Theta\,.
\end{equation}
The 'Wigner crystal' slides if scaling dimension, $\Delta$, of $J=2$ processes below $3/2$. Otherwise, when scaling dimension,
\begin{equation}
\Delta=\frac{4\,K_{\parallel}}{N}\leq\frac{3}{2}\,,
\end{equation}
multi-backscattering processes pin  CDW  \cite{Gim}.

{\it Spontaneous TRS breaking} --
The pinning of Wigner crystal structure is always accompanied by TRS breaking. The expectation of terms like $\cos({\bf j}{\bm\phi})$ must vanish in TRS system if corresponding vector ${\bf j}$ belongs to the sector of odd integer $J={\bf q}{\bf e}$. Freezing all $N$ fields ${\bm\phi}$ leads to all such terms acquiring finite value that means spontaneous TRS breaking.

Note that we have not referred to the Haldane criterion \cite{hal} since we are dealing with weak interaction problem and, therefore, do not assume that all amplitudes of all allowed processes are infinitely strong and open gaps. Our choice of interactions was motivated by RG analyses and only those terms that are RG-relevant became potential candidate for opening of gaps. It turned out that there are exactly $(N-1)$ such terms and they do not break TRS. All these terms contain only density fields ${\bm\phi}$ and, therefore, commute with each other, making check of Haldane compatibility condition \cite{hal} unnecessary. An additional term that potentially could be relevant for closing the remaining gap, should also contain density field since all current fields are irrelevant. When this additional $J\neq 0$-term becomes relevant, it necessarily breaks TRS and this fact is not related to the parity of the KDs number (similar to fractional topological insulator \cite{ster}).

Let us comment here on the correspondence between our week interaction problem and strong interaction problem analysed in \cite{neu}. As it was shown above, the repulsion implies RG relevance of the terms containing density ${\bm\phi}$-fields only and, therefore, one immediately constructs $(N-1)$ vertex operators with $J={\bf j}{\bf e}=0$ since the subspace of the vectors ${\bf j}$ orthogonal to the vector ${\bf e}$ is $(N-1)$-dimensional. Disorder allows vertices with $J\neq 0$ which may gap the last mode exhausting $N$-dimensional space of vectors ${\bf j}$. Our construction is dictated by RG analysis and leaves us no choice. Had we dealt with a strong interaction problem and used $(N-1)$ current ${\bm\theta}$-fields instead (like it was done in \cite{neu}), we would immediately present $(N-1)$ interaction vertices with $Q={\bf q}{\bf e}=0$. But the extra term that could potentially gap the remaining mode could not be picked up from the same current fields due to neutrality $Q=0$ condition. This vortices neutrality condition breaks the duality between density and current vertices. The extra term could come from invoking conjugate density field and that is where the Haldane criterion \cite{hal} becomes crucial in justification that the additional term is consistent with already built $(N-1)$ vertices. One might think that this situation may appear in weak interaction problem when superconducting vertices become relevant perturbations at $K_{\perp}>1$ but it is not obvious because describing superconductivity one has to include anomalous terms that break $Q=0$ neutrality condition.

{\it Phase diagram} -- In general, the phase diagram should be drawn in three-dimensional space of parameters characterising intra-channel interaction (standard Luttinger parameter $K$) and two (density-density and current-current) inter-wire interactions $\alpha_{\pm}$. Below we will analyse in detail the commonly accepted model that includes only density-density interaction assuming current-current interactions matrix $\hat{V}_-=\hat{\mathbb{1}}$ in Eq.~\eqref{V}. The two parameters $K$ and $\alpha_+$  characterise intra- and inter-mode interactions, respectively, and define the effective Luttinger parameters,
\begin{eqnarray}
K_{\perp}&=&K\,(1-\alpha_+)^{-1/2}\,,\\
K_{\parallel}&=&K\,\left[1+(N-1)\alpha_+\right]^{-1/2}\,,
\end{eqnarray}
The region of existence of a delocalised (conducting) mode of repulsive electrons, $K < 1$, in CDW regime, $K_{\perp} < 1$, and irrelevant pinning by disorder, $\Delta > 3/2$, is defined by the inequality:
\begin{equation}\label{phase}
\frac{3N}{8}\left[1+(N-1)\alpha_+\right]^{1/2} < K < (1-\alpha_+)^{1/2}\,.
\end{equation}

\begin{figure}
\includegraphics[width=0.9 \linewidth]{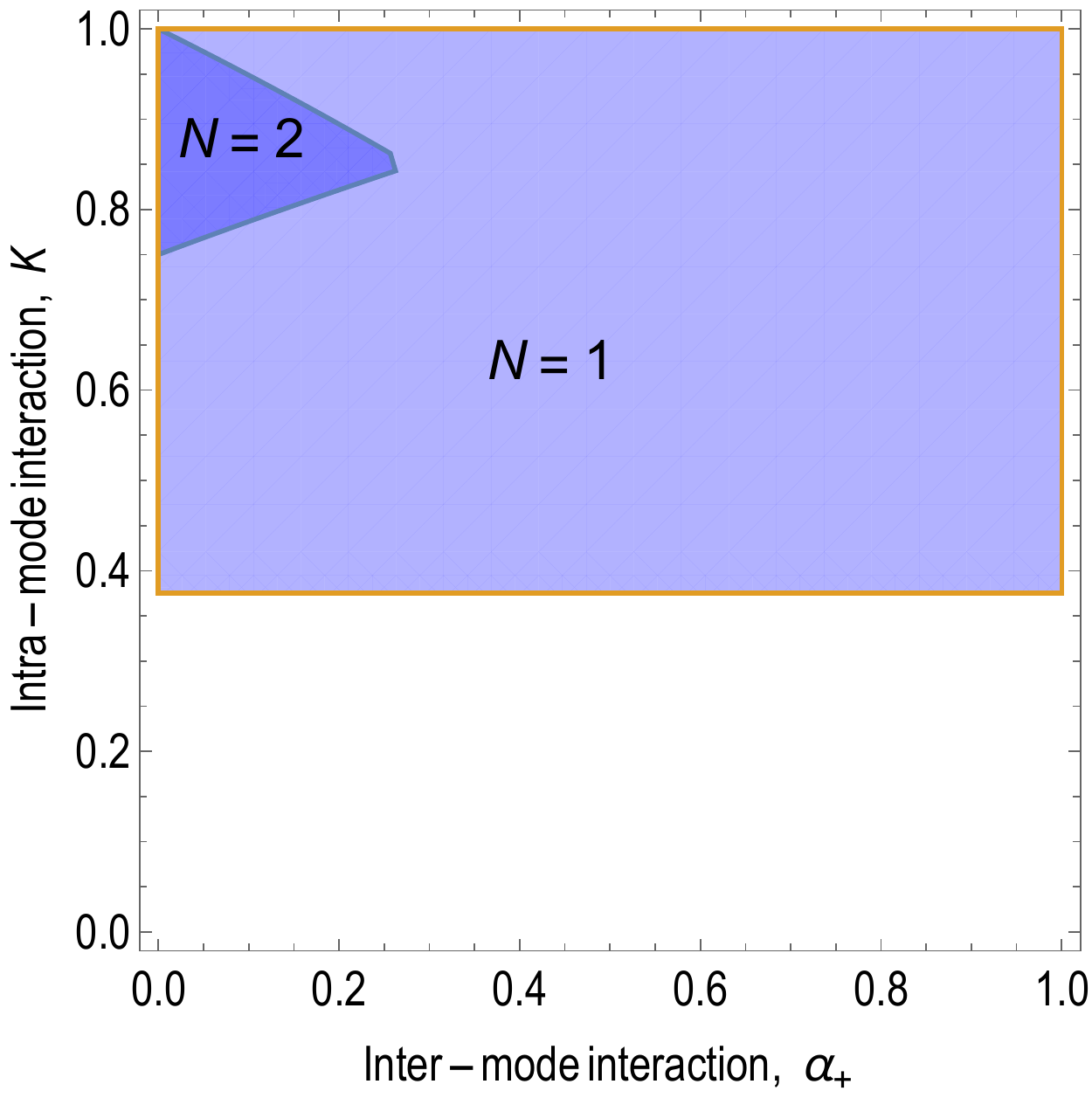}
\caption{The phase diagram for a set of $N$ Kramers doublets under repulsive density-density interaction. The only two stable states, $N=1$ and $N=2$, are shown by light and dark blue regions correspondingly.}
\label{CDW_two}
\end{figure}

It is clear from these inequalities that {\it more than two ($N> 2$) interacting Kramers doublets are always pinned by disorder}. In Fig.~\ref{CDW_two} we show regions of stability for systems with $N=1$ and $N=2$ KDs. Both regions are defined by the inequalities Eq.~(\ref{phase}). In the single KD situation there is no inter-channel interaction and one should put $\alpha_+=0$ in the inequalities Eq.~\eqref{phase} for $N=1$.

A single conducting state even for a system with a pair of KDs ($N=2$) survives pinning only in a small region of interaction parameters. It exists for weak interactions and immediately disappears if either inter- or intra-mode interaction becomes strong. There is no solution to the inequality \eqref{phase} for $N>2$ meaning that higher number of KDs is unobservable since the system becomes insulating for any inter- and intra-interaction strength due to the pinning by disorder.

{\it Conclusions} -- 
We have studied a topological insulator with $N$ Kramers doublets at the edge in the model of 'Coulomb blockade' type (long range featureless) interaction. This type of interaction is relevant for the situation when the screening radius is much larger than the Fermi wavelength (i.e. the width of the region occupied by edge states). We have shown that in a clean system the perturbations allowed by TRS always open $N-1$ gaps. In a non-superconducting regime, when the relevant perturbation in a clean system is of CDW type, the opening of $N-1$ gaps reduces the number of conducting edge channels to one. The disorder can either reduce the number of conducting channels to zero or leave the {\it only} conducting channel unaffected. We have found that only single Kramers doublet or a pair of them may survive pinning by disorder. The  phase diagram contains a small pocket where both $N=1$ and $N=2$ are conducting. The relatively small size of this region might be responsible for elusiveness of experimental observation of states with two Kramers doublets.  Any higher number $N>2$ of Kramers doublets, irrespective of parity, gets fully localised by disorder when density-density repulsion is taken into account. This conclusion comes from the fact that a featureless long range interaction between Kramers doublets in a topological insulator leads to formation of a {\it single} gapless edge mode that get easily pinned by the disorder-induced two-particle backscattering.

{\it Acknowledgments} -- 
This work was supported by the Leverhulme Trust Grant No.\ RPG-2016-044 (IVY). The authors are grateful for hospitality extended to them at  the Center for Theoretical Physics of Complex Systems, Daejeon, South Korea.
%\begin{acknowledgements}

\end{document}